 \documentclass[final,5p,times,twocolumn]{elsarticle}

\usepackage{amssymb}

\usepackage{balance}
\usepackage{ulem}

\usepackage[usenames]{color}

\newcommand{\cred}{ }          

\usepackage{lineno,hyperref}

\journal{Solar Energy Materials and Solar Cells}

\begin{document}

\begin{frontmatter}

\title{Dynamic electrical behavior of halide perovskite based solar cells}

\author[1,2]{G. A. Nemnes\corref{mycorrespondingauthor}}
\cortext[mycorrespondingauthor]{Corresponding author. Tel.: +40 (0)21 457 4949/157. \\ {\it E-mail address:} nemnes@solid.fizica.unibuc.ro (G.A. Nemnes).}
\address[1]{University of Bucharest, Faculty of Physics, Materials and Devices for Electronics and Optoelectronics Research Center,\\ 077125 Magurele-Ilfov, Romania}
\address[2]{Horia Hulubei National Institute for Physics and Nuclear Engineering, 077126 Magurele-Ilfov, Romania}

\author[3]{Cristina Besleaga}

\author[3]{A. G. Tomulescu}

\author[3]{Ioana Pintilie}

\author[3]{L. Pintilie}
\address[3]{National Institute of Materials Physics, Magurele 077125, Ilfov, Romania}

\author[4]{K. Torfason}

\author[4]{A. Manolescu}
\address[4]{School of Science and Engineering,
        Reykjavik University, Menntavegur 1, IS-101 Reykjavik, Iceland}




\begin{abstract}
A dynamic electrical model is introduced to investigate the hysteretic effects in the J-V characteristics of
perovskite based solar cells. By making a simple ansatz for the polarization relaxation, our model is able to reproduce qualitatively and quantitatively detailed features of measured J-V characteristics. Pre-poling effects are discussed, pointing out the differences between initially over- and under-polarized samples. In particular, the presence of the current overshoot observed in the reverse characteristics is correlated with the solar cell pre-conditioning. Furthermore, the dynamic hysteresis is analyzed with respect to changing the bias scan rate, the obtained results being consistent with experimentally reported data: the hysteresis amplitude is maximum at intermediate scan rates, while at very slow and very fast ones it becomes negligible. The effects induced by different relaxation time scales are assessed.
The proposed dynamic electrical model offers a comprehensive view of the solar cell operation, being a practical tool for future calibration of tentative microscopic descriptions. 
\end{abstract}


\end{frontmatter}


\section{Introduction}

The rapid increase of power conversion efficiencies (PCEs) in hybrid organic-inorganic solar cells based on 
halide perovskites received a lot of attention recently \cite{stranks,park,chen1,zhao,song}.  
Hybrid lead halides are regarded as one of the most 
promising class of materials for photovoltaic applications, due to their 
\cred{high optical absorption coefficient} \cite{kim1}, large
diffusion lengths for both electrons and holes \cite{xing}, while the band gap may be tuned by changing the halogen composition towards an optimum range \cite{eperon,walsh1}.
Moreover, halide perovskites layers may be obtained by low temperature 
solution processing techniques, which make them suitable for integration in flexible plastic substrates \cite{lili}.  

While the high PCEs and potentially low production costs are important assets, a number of challenges 
and open questions still need to be addressed. These concern especially the solar cell stability under ambient conditions \cite{giu}, a better understanding of the device operation, and the optimization of the perovskite/substrate interface \cite{nemnes,nicolaev}. Furthermore, the J-V characteristics typically display hysteresis when measured under forward (short-circuit to open-circuit bias) 
and reverse bias scans \cite{snaith,dualeh,sanchez,wei,tress,chen3}. This behavior is usually influenced by the solar cell pre-conditioning and the bias sweeping rate (BSR). 
The hysteretic effects pose a significant challenge for determining the PCE of the solar cell, but they may also provide information regarding transport and polarization mechanisms.    

\cred{
The hysteresis of the dynamic J-V characteristic is also known for other
types of solar cells, but only when the voltage sweep is fast \cite{koide},
whereas in perovskite cells it is obtained at sweep rates of Volt per
minute.  The origin of this anomalous hysteresis} is still under debate and
has been attributed to different phenomena, such as a giant photoinduced
dielectric response \cite{perez}, ferroelectric effects \cite{wei,chen4,frost},
\cred{ion migration \cite{tress,meloni} possibly accompanied by charge trapping
and de-trapping \cite{snaith}, or unbalanced distributions of electrons
and holes \cite{xing,bergmann,heo}.}

It is a well known fact that variations between the J-V characteristics under forward and reverse bias scans arise in the presence of a large capacitance, for a high BSR.
In forward scan, a fraction of the photo-generated current charges the built-in capacitor, leading to smaller values in the current than in reverse scan, when the capacitor releases the charges, leading to relatively larger PCEs. 
In perovskite solar cells a slow decay of the photovoltage was observed and initially related to capacitive effects introduced by a giant photoinduced dielectric constant \cite{perez}.
Ferroelectricity is another potential cause considered for the observed hysteresis. Yet it is a controversial hypothesis, as the time-scales involved are typically below ms \cite{leguy}, which lends the idea of microscopic ferroelectricity \cite{charles}. 
It is important to note, however, that capacitive effects alone cannot reproduce different pre-conditioning effects induced by initial light soaking and pre-poling by different biases \cite{chen3}.

As was later found, the capacitive effects are quite independent on the thickness of the perovskite layer, which suggests that the induced polarization is related to interfacial properties \cite{almora}, including ion accumulations near the electrodes. Trapping and de-trapping processes near the interfaces and grain boundaries are also thought to play a role in the hysteretic behavior \cite{snaith}.  
 
More recently the ionic migration mechanism is gaining ground. The perovskite being known as a ionic conductor, Tress {\it et al.} \cite{tress} argued for a non-electronic reason for the hysteresis, based on several observations, such as: relaxation time scales between seconds to minutes, strong dependence on BSR and temperature, or counterbalancing of the polarization for applied voltages of relatively high negative values. A recently proposed drift-diffusion model taking into account ionic migration however shows that the observed hysteretic behavior may be obtained only by including electronic charge traps, serving as recombination centers \cite{reenen}.   

The paper is structured as follows. In Section \ref{model}, the dynamic electrical model is introduced in form of an equivalent circuit model specified by a given {\it ansatz} for the polarization relaxation, which is calibrated on measured J-V characteristics. A brief description of the experimental device structures and methods is also indicated. Section \ref{results} presents simulated J-V characteristics compared to experimental data, describing the hysteretic effects, which are influenced by solar cell preconditioning and bias scan rate. The analysis is focused on the variation of the polarization in time, assuming a certain relaxation time scale. Furthermore the influence of different relaxation time scales is investigated. Conclusions are collected in Section \ref{conclusions}.

\section{Model and Methods}
\label{model}

\subsection{The standard dynamic electrical model}

\begin{figure}[t]
\centering
\includegraphics[width=8.cm]{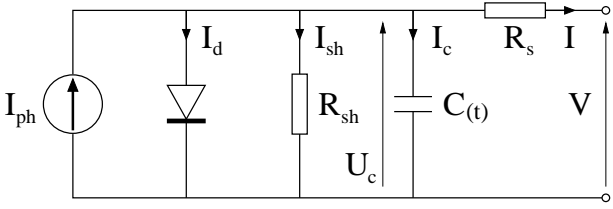}
\caption{Dynamic electrical model of the solar cell. The basic elements are: the current source, a diode with ideality factor $n$, the shunt and series resistances and the capacitance, which is here introduced by a polarization mechanism. 
}
\label{DEM}
\end{figure}

In the standard dynamic electrical model the solar cell is described by the equivalent electric circuit 
depicted in Fig.\ \ref{DEM}.
The common elements also found in the static model are: \cred{the current source, 
i. e. the photocurrent $I_{ph}$, defined by the number of electron-hole pairs generated per time unit}; 
the diode, with the ideality factor $n$; the shunt resistance $R_{sh}$, which accounts for the recombination processes; the series resistance
$R_s$, representing the resistance of the active layers and the contacts.
In addition to the static model, a parallel capacitance $C$ is present, which is generally responsible for 
dynamically non-equivalent forward and reverse bias scans. 

The \cred{current-voltage} characteristics of the solar cell under illumination, written for the first quadrant ($I>0$), is found from :
\begin{equation}
I_{ph} = I_d + I_{sh} + I_{c} + I,
\label{IVeq}
\end{equation}
where \cred{$I$ is the current in the external circuit at applied bias $V$ and}
\begin{eqnarray}
I_d &=& I_s \left( e^{\frac{q(V + I R_s)}{n k_B T}}-1\right) \ , \\
I_{sh} &=& \frac{V + I R_s}{R_{sh}}, \\
\label{Ic}
I_{c} &=& C\frac{\partial(V + I R_s)}{\partial t} + (V + I R_s)\frac{\partial C}{\partial t} \ , 
\end{eqnarray}
where $I_s$ is the reverse saturation current of the diode with ideality factor $n$ 
($n=1$ for an ideal diode), \cred{and $k_B$ is Boltzmann's constant}. 
In general we assume the quantities $I=I(t)$, $V=V(t)$ and $C=C(t)$ are time-dependent.

One obtains the following first order differential equation for the $I(t)$:
\begin{eqnarray}
\label{Idiffeq1}
-R_s C \frac{\partial I}{\partial t} &=&
      I_s \left( e^{\frac{q(V + I R_s)}{n k_B T}}-1\right) \nonumber \\
&+& \left(\frac{R_s}{R_{sh}} + R_s \frac{\partial C}{\partial t} + 1 \right) I \nonumber \\
&+& \frac{V}{R_{sh}} + C \frac{\partial V}{\partial t}
    + V\frac{\partial C}{\partial t} - I_{ph} \ .
\end{eqnarray}
Assuming a certain initial condition $I(t=0)$ one can find the time-dependent
behavior of the current under forward and reverse bias sweeps introduced by $V(t)$. 

In the case of a negligible capacitance $(C=0)$ one recovers the characteristic equation of the static model:
\begin{equation}
\left(\frac{R_s}{R_{sh}} + 1 \right) I + I_s \left( e^{\frac{q(V + I R_s)}{n k_B T}}-1\right) 
	+ \frac{V}{R_{sh}} - I_{ph} = 0 \ . 
\end{equation}

\subsection{Modeling the capacitor current}

More generally we may express the displacement current $I_c$ in equation (\ref{Ic}) as:
\begin{equation}
I_{c} = {\mathcal A} \ \frac{\partial D(t)}{\partial t}
      = {\mathcal A} \ \epsilon_0 \ \frac{\partial E(t)}{\partial t} + {\mathcal A} \ \frac{\partial P(t)}{\partial t} \ ,
\end{equation} 
where $D(t) = \epsilon_0 E(t) + P(t)$, \cred{$\epsilon_0$ being the vacuum permittivity},
and ${\mathcal A}$ is the device area.
Similarly to the interface polarization model proposed in Ref.\ \cite{pintilie} 
we may further split $P(t)$ into a linear term 
\cred{with respect to the electric field,
$P_l(t)= \epsilon_0 \chi_l E(t)$, and a non-linear contribution arising from various 
complex polarization mechanisms in the active medium, $P_{nl}(t)$.  Indeed, the 
nonlinear polarization is a result of a nonlinear response of the charge to the combined effect of
light and electric field inside the device or at the interfaces with the electrodes.
The microscopic origin of it is however beyond the scope of our paper.}

Assuming the planar capacitor has a typical distance $d$ between the plates, 
\cred{we reformulate} Eq.\ (\ref{Ic}) as
\begin{equation}
I_{c} = C_0 \ \frac{\partial U_c(t)}{\partial t} + {\mathcal A} \ \frac{\partial P_{nl}(t)}{\partial t} \ ,
\label{IcP}
\end{equation}
where $C_0=\epsilon {\mathcal A}/d$ 
is a constant geometrical capacitance, which captures the linear polarization 
via the dielectric constant of the material $\epsilon = (1+\chi_l) \epsilon_0$, 
and $U_c=V + I R_s$ is the bias on the capacitor. For the two terms in Eq.\ (\ref{IcP}) we introduce the notations 
$I_c^{(l)}=C_0 \partial U_c(t)/\partial t$ and $I_c^{(nl)}= {\mathcal A} \times \partial P_{nl}(t)/\partial t$. 
 
Eq. (\ref{Idiffeq1}) then becomes:
\begin{eqnarray}
\label{Idiffeq2}
-R_s C_0 \frac{\partial I}{\partial t} &=&
      I_s \left( e^{\frac{q(V + I R_s)}{n k_B T}}-1\right) \nonumber \\
&+& \left(\frac{R_s}{R_{sh}} + 1 \right) I \nonumber \\
&+& \frac{V}{R_{sh}} + C_0 \frac{\partial V}{\partial t}
    + {\mathcal A}\frac{\partial P_{nl}}{\partial t} - I_{ph}. 
\end{eqnarray}

The current-voltage characteristics of the solar cell under illumination may be found from Eq. (\ref{Idiffeq2}) provided the variation of the polarization in time is known. Depending on the relaxation processes involved different models for $\partial P_{nl}/\partial t$ may be introduced.




As a first {\it ansatz}, in the single relaxation time approximation, we assume the following behavior for the polarization $P_{nl}(t)$:
\begin{equation}
\frac{\partial P_{nl}}{\partial t} = \frac{P_{nl,\infty}(U_c(t)) - P_{nl}(t)}{\tau},
\label{dPdt}
\end{equation}
where $P_{nl,\infty}$ is not constant, but rather a function of applied bias on the capacitor $U_c$. 
We take $P_{nl,\infty} = (U_c/V_{oc}) P_\infty$, where $P_\infty$ is a system dependent quantity, as well as the relaxation time $\tau$, and $V_{oc}$ is the open circuit bias. 
\cred{Several polarization mechanisms are compatible with our approach, which may have the origins in the displacement of the electronic and ionic components inducing Maxwell-Wagner-Sillars type polarization in an inhomogeneous medium, alignment of microscopic ferro-electric domains, charged defect migration or a combination of these. Since rather slow relaxation times have been observed, this suggest a glassy type behavior and possibly multiple relaxation time scales need to be considered \cite{tress,roiati,lee}. However the microscopic origins of the polarization are beyond the scope of this paper and will be considered elsewhere.}

The equations (\ref{Idiffeq2}) and (\ref{dPdt}) form a system of coupled differential equations, which give the time dependent current, $I(t)$, and the non-linear polarization, $P_{nl}(t)$, in the device, starting from the initial conditions, $I(t=0)=I_0$ and $P_{nl}(t=0)=P_0$. Being set by the solar cell preconditioning, $I_0$ and $P_0$ are important parameters in establishing the dynamic current-voltage characteristics.

\begin{figure}[t]
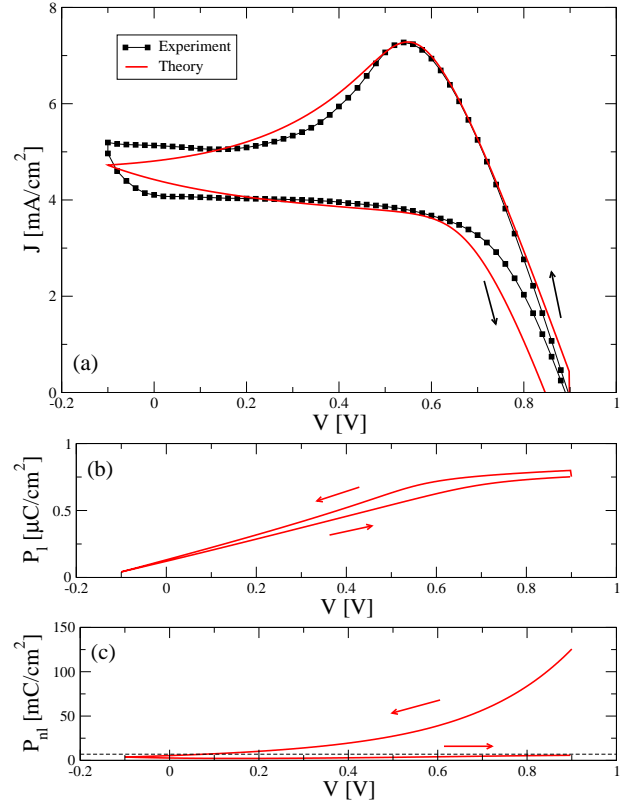

\centering
\includegraphics[width=8.cm]{figure2a}\\
\includegraphics[width=8.cm]{figure2b}\\
\includegraphics[width=8.cm]{figure2c}
\caption{{(a)} Reverse-forward bias scan of a typical pre-poled sample: calculated \cred{J-V} characteristics (red lines) in comparison with experimental data (black points). The forward and reverse data corresponds to increasing and decreasing voltage, respectively. The initial pre-set voltage is $V_0=1.6$ V. {(b) and (c)} show the polarization as a function of applied bias, for both linear an non-linear terms, $P_l$ and $P_{nl}$, respectively. The dotted horizontal line marks \cred{$P_\infty=6.97$ mC/cm$^2$}.
}
\label{I_P}
\end{figure}

\subsection{Experimental device structures and methods}

The CH$_3$NH$_3$PbI$_{3-x}$Cl$_x$ based solar cells were fabricated using two-step deposition method 
\cred{\cite{burschka,xiao,jiang}}: the PbI$_2$ + PbCl$_2$ films were spin-coated on the FTO/TiO$_2$/mp-TiO$_2$ structures and then dried for
15 minutes at 110$^{\circ}$C. In order to obtain the CH$_3$NH$_3$PbI$_{3-x}$Cl$_x$  hybrid compound, the FTO/TiO$_2$/mp-TiO$_2$/PbI$_2$+PbCl$_2$
 structure was dipped in a solution of CH$_3$NH$_3$I and anhydrous 2-propanol
{(10mg$^{-1}$ml$^{-1}$)} and then rinsed in 2-propanol.
The  hole   transporter  material -- 2,2',7,7'-Tetrakis-(N,N-di-4-methoxyphenylamino)-9,9'- spirobifluorene (spiro-OMeTAD) {dissolved in chlorobenzene} and tert-butylpyridine, litiumbis (trifluoromethylsulphonyl) imide in acetonitrile -- was deposited by spin-coating in N$_2$
  enriched
atmosphere, at room temperature ($\sim$24$^{\circ}$C) and humidity less than 10\%. 
\cred{Gold electrodes with area 0.215 cm$^2$ were deposited as top contacts,
while the illuminated area is 0.125 cm$^2$}.
\cred{The photo-electrical measurements were performed at room temperature, 
in air, under AM 1.5  conditions with a computer controlled setup 
consisting of a Solar Simulator (Newport-Oriel 170 VeraSol LED Controller)
 and a Keithley 2601B Source Meter. The VeraSol LED solar simulator ensures 
stable power during the operating time. The electrical measurement 
were performed using Sample \& Hold method.}

\section{Results and discussion}
\label{results}

We discuss in the following dynamic J-V characteristics with hysteresis effects which arise due to pre-polation and BSR, and we detail the changes introduced by different relaxation time scales. 

\begin{figure}[t]
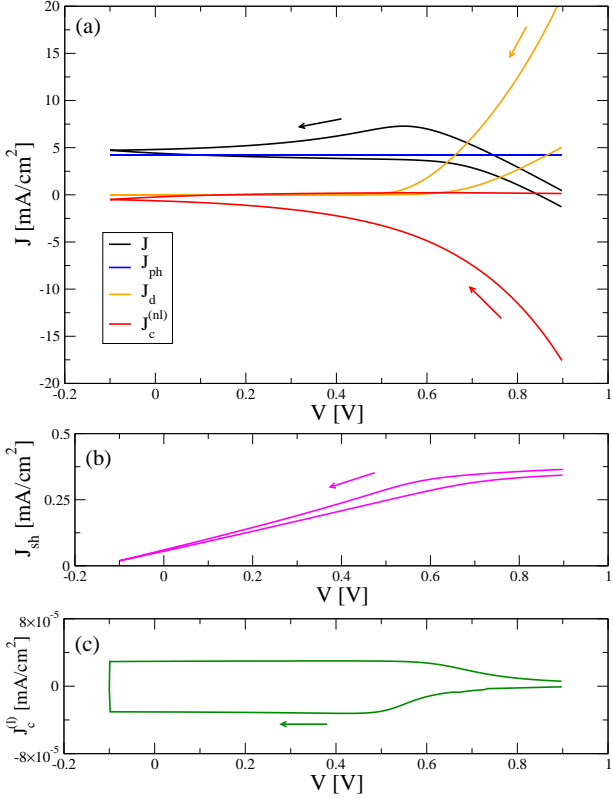

\centering
\includegraphics[width=8.cm]{figure3a}\\
\includegraphics[width=8.cm]{figure3b}\\
\includegraphics[width=8.cm]{figure3c}
\caption{Decomposition of the total current in the solar cell under illumination according to Eq.\ (\ref{IVeq}). \cred{The most significant currents, $J_{ph}$, $J_d$ and $J_{c}^{(nl)}$, as well as the current in the external circuit $J$ are presented in (a), while the smaller currents $J_{sh}$ and $J_{c}^{(l)}$ are shown in (b) and (c), respectively.} 
}
\label{I_decomposition}
\end{figure}

\subsection{Pre-poling effects}

\label{pre-poling}

The test case parameters of the investigated solar cells are calibrated based on the experimental data and device structure. We start our analysis with a typical J-V characteristics of a reverse-forward bias scan, indicated in Fig.\ \ref{I_P}, which is reproduced considering the following model elements:
$R_s=250$ $\Omega$, $R_{sh}=20$ k$\Omega$, $n=1.47$, \cred{$I_s=0.1$ pA,} $I_{ph}=0.53$ mA; \cred{$C_0=0.19$ $\mu$F} is the geometrical capacitance of a planar capacitor, with \cred{the active region area ${\mathcal A}=0.215$ cm$^2$}, the distance between the plates $d=100$ nm and the relative dielectric permittivity $\epsilon_r=100$; the steady state polarization at open circuit voltage is \cred{$P_\infty(V_{oc})=6.97$ mC/cm$^2$,} which is within the range reported in Ref. \cite{tress} and the initial polarization is $P_0=18 P_\infty$; the open circuit voltage is $V_{oc}=0.9$ V; a characteristic relaxation time $\tau=11.5$ s is assumed, \cred{consistent with reported data of transient current-voltage measurements \cite{unger}}; a constant bias scan rate $\alpha=\partial V/\partial t= 20$ mV/s is considered, as in the experimental measurements. \cred{In the following we shall refer to current densities $J=I/{\mathcal A}_{i}$, where ${\mathcal A}_{i}=0.125$ cm$^2$ is the illuminated area.}

Figure\ {\ref{I_P}(a)} is illustrative for a pre-poled sample, typically induced by pre-setting the initial bias $V_0=1.6$ V, larger than $V_{oc}$ \cite{wei,chen2}, i.e. reversing the sign of the current relatively to the normal operational mode of the solar cell. The J-V characteristics presents dynamic hysteresis, with a pronounced maximum in the reverse scan. As we shall see in more detail in the subsequent analysis, the magnitude of this overshoot is directly related to the initial polarization $P_0$ and depends also on the characteristic relaxation time $\tau$. By assuming a pre-poled sample with $P_0 > P_\infty$ and lowering the bias in the reverse scan the extra-current due to polarization relaxation is pumped out in the external circuit, which accounts for the bump in the current. One should note that this situation occurs due to the solar cell pre-conditioning and it is not simply an effect of capacitance induced dynamic hysteresis. Moreover, the overshoot effect is reproduced experimentally by pre-setting the initial bias larger than $V_{oc}$ in dark conditions, and therefore it seems independent on the illumination in the pre-setting time interval. 
\cred{One should mention that the solar cells typically exhibit lower performance over time, due to aging as well as potential damage due to electrical stress (see Supplemental material). The PCE typically drops more than a half in 600 hours, from $\sim$11\% (fresh cells) to $\sim$4\%, followed by a much slower degradation observed up to 1000 hours. In order
to diminish potential effects due aging, we performed the dynamic J-V experiments on this apparent stability plateau.} 

\begin{figure}[t]
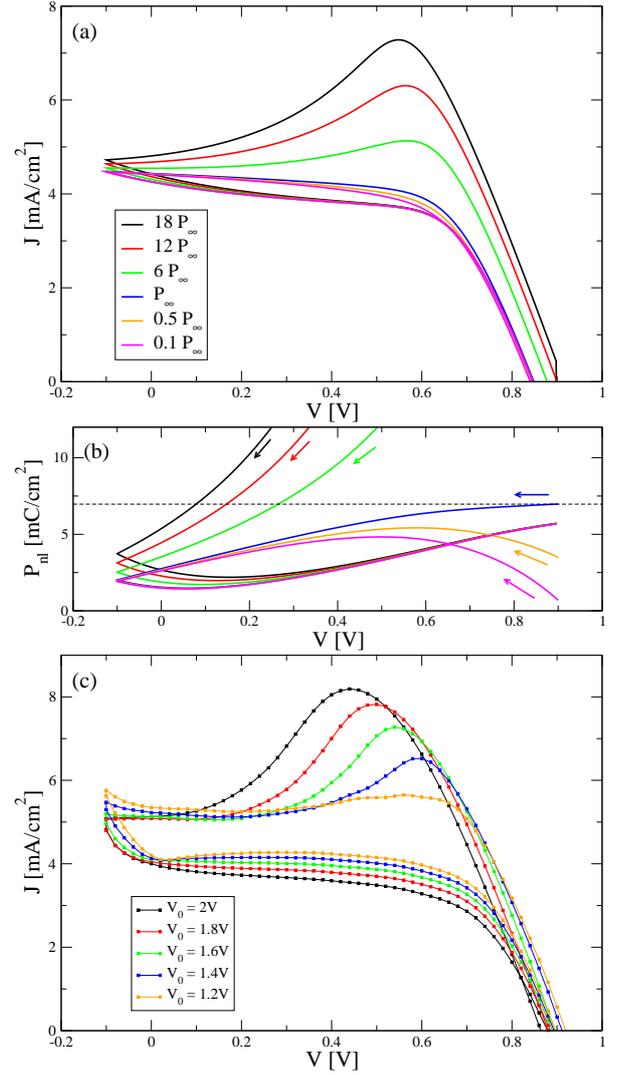

\centering
\includegraphics[width=7.8cm]{figure4a} \\
\includegraphics[width=7.8cm]{figure4b} \\
\includegraphics[width=7.8cm]{figure4c}
\caption{Reverse-forward bias scans for different pre-poling conditions, defined by the initial 
polarization $P_0 = 0.1, 0.5, 1, 6, 12, 18 P_\infty$: (a) the J-V characteristics, (b) the behavior of the polarization, with the focus on the values below $P_\infty$. 
(c) experimental J-V characteristics, with the samples pre-poled by setting the initial bias $V_0$ = 2, 1.8, 1.6, 1.4, 1.2 V. 
}
\label{I_P_different_P0}
\end{figure}

Furthermore, as one can see from {Figs.\ \ref{I_P}(b) and (c)}, the J-V characteristics can be correlated with the polarization variation in time. First one should note the rather different magnitudes of the two polarization components, $P_l \ll P_{nl}$, since the geometrical capacitance is rather small, in the $\mu$F range. Secondly $P_{nl}$ decreases significantly as the applied voltage moves towards negative values, while on forward bias the polarization starts growing, albeit with a smaller rate towards $P_\infty$. Both the rate of the decrease in reverse and the slight increase in forward regime are set by \cred{the bias scan rate and} the single relaxation time $\tau$ assumed.

For a better understanding of the role of the different elements in the DEM, we plotted in Fig.\ \ref{I_decomposition} the individual \cred{current densities, $J_{ph}$, $J_d$, $J_{sh}$, $J_{c}^{(l)}$ and $J_{c}^{(nl)}$,} which give the current $I$ in the external circuit. While the photogenerated current $I_{ph}$ is assumed independent on the applied bias, the other four currents depend on the operating point, specifically on $I$ at a certain bias $V$, which is determined from the coupled system of equations (\ref{Idiffeq2}) and (\ref{dPdt}). Some comments on the specific behavior of the partial currents are worth to be mentioned. The diode current $I_d$ is significantly different in reverse and forward scans, due to the exponential dependency on the relatively small differences introduced by the $V+IR_s$ term. However, the current through the shunt resistance $I_{sh}$ is directly proportional with this term and the differences between reverse and forward scans are correspondingly small. This may be explained as follows.
One may see that $I R_s$ is at most $\approx0.22$ V, when considering the maximum current $I\approx0.9$ mA (in reverse), at $V\approx0.55$ V, and the $I R_s$ bias differences between forward and reverse currents are even smaller. At larger biases, $V>0.55$ V, these bias differences become negligible compared to $V$. On the other hand, at smaller biases, $V\approx0$, given the shape of the dynamic J-V characteristics, the $I R_s$ is quite similar for both forward and reverse scans. Therefore, overall, $I_{sh}$ is quite similar in reverse and forward scans. The current through the capacitor $I_c$ has two components, $I_{c}^{(l)}$ and $I_{c}^{(nl)}$, corresponding to the geometrical capacitance and non-linear polarization, respectively. 
The current $I_{c}^{(l)}=C_0 (\partial V/\partial t) + C_0 R_s (\partial I/\partial t)$ is rather small, due to the small geometrical capacitance $C_0$ and moderate bias scan rate $\alpha$ considered, having small variations around $V_{oc}$, where $\partial I/\partial t$ is the largest. Therefore, in this context, $I_{c}^{(l)}$ may be completely neglected. 
\cred{Instead, $I_{c}^{(nl)}$ has a large 
contribution during the reverse scan. The polarization is decreasing,
reaching values below $P_\infty$, while $I_{c}^{(nl)}$ is negative 
(compared to the photocurrent), and thus the total current $I$ 
in the external circuit increases. On the forward scan, the polarization
slowly increases towards $P_\infty$, yielding a rather small positive
$I_{c}^{(nl)}$, which lowers the current $I$.}

We next investigate the effects of different pre-poling conditions. Figure\ \ref{I_P_different_P0}(a) shows
the changes introduced in the J-V characteristics by varying the initial polarization $P_0$, while the other parameters remain the same. One may describe two operating regimes: an over-polarization regime, for $P_0>P_\infty$, \cred{obtained by setting the starting bias larger than $V_{oc}$} and an under-polarization regime, for $P_0<P_\infty$, \cred{which may be obtained by pre-setting the initial bias smaller than $V_{oc}$}.
The over-polarized samples exhibit large dynamic hysteresis, with the overshoot being diminished as the $P_0$ decreases. In the under-polarization regime, the J-V characteristics of reverse and forward scans are almost overlapping with a crossing point moving towards lower biases as $P_0$ is decreased, e.g. $\sim$0.75 V for $P_0 = 0.5 P_\infty$, $\sim$0.65 V for $P_0 = 0.1 P_\infty$. 
This feature may be directly correlated with
\begin{figure}[h]
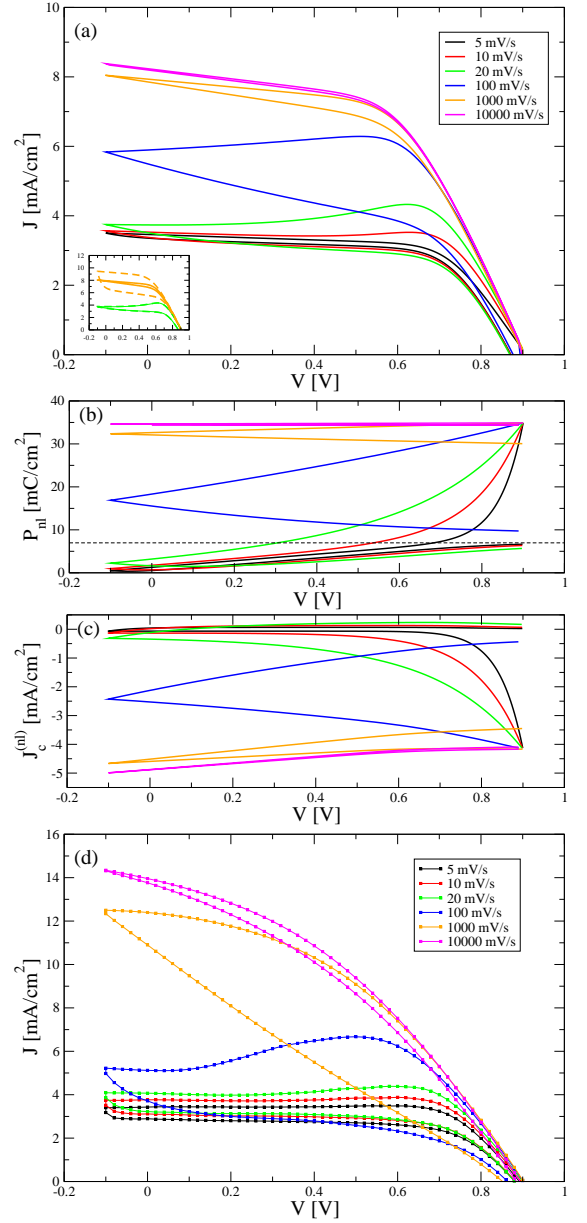

\centering
\includegraphics[width=7.3cm]{figure5a} \\
\includegraphics[width=7.3cm]{figure5b} \\
\includegraphics[width=7.3cm]{figure5c} \\
\includegraphics[width=7.3cm]{figure5d} 
\caption{Dynamic hysteresis under different bias scan rates: (a) simulated J-V characteristics, qualitatively similar with our experimental curves and other data as well \cite{tress}; the inset shows simulated J-V characteristics, taking into account a large value for the dielectric constant, $\epsilon_r=10^5$ (dashed lines), as reported in Ref.\ \cite{perez}; (b) the corresponding polarization; \cred{(c) the current component $J_{c}^{(nl)}$; (d)} experimental J-V characteristics, measured at different BSRs.}
\label{IP_different_BSR}
\end{figure}
 the similar behavior of the polarization represented in Fig.\ \ref{I_P_different_P0}(b). It is worth noting that for the two regimes we have $(\partial P/\partial V)(V_{oc}) > 0$ for $P_0>P_\infty$ and $(\partial P/\partial V)(V_{oc}) < 0$ for $P_0<P_\infty$, while for the separator characteristics $P_0=P_\infty$ we have $(\partial P/\partial V)(V_{oc}) = 0$. 
Experimental data indicated in Fig.\ \ref{I_P_different_P0}(c) shows a similar behavior. The sample is pre-poled by an initial bias $V_0 > V_{oc}$ and subsequently measurements are performed keeping the BSR at 20 mV/s. The J-V characteristics correspond to the over-polarization regime, where the current overshoot becomes more and more prominent as $V_0$ is increased, which induces larger $P_0$. The maxima are also shifted towards smaller biases, an effect which may be correlated with a change in the relaxation time, as discussed in more detail in the last section.

\subsection{Bias scan rate effects}

In order to further analyze the dynamic hysteresis effects and test the validity of our model, we explore the J-V characteristics at different bias scan rates.  Several reverse-forward scans with different BSRs are presented in Fig.\ \ref{IP_different_BSR}(a). The widest hysteresis occurs at moderate scan rates, while at very slow and very fast rates it is diminished. 
This picture is explained as follows. At lower scan rates, the capacitive effects are less important, while at very high rates the relaxation cannot follow the rapidly changing fields.

This is consistent with our model and, in particular, with the polarization behavior depicted in Fig.\ \ref{IP_different_BSR}(b). At small rates, $\alpha= 10 - 20$ mV/s, there is a rather sharp decrease in polarization referred to the applied bias, as the scan is performed in rather long time interval. In forward scan, a slight increase of the polarization may be observed, below $P_\infty$. Increasing the scan rate to 100 mV/s, taking into account that $\tau=11.5$ s there is no sufficient time for polarization to drop below $P_\infty$. At even higher rates the decrease of the polarization, \cred{$\partial P_{nl}/\partial V$}, becomes smaller and, in the limiting case of $\alpha=10000$ mV/s, the polarization remains almost constant.
\cred{However, the variation in time of the polarization, $\partial P_{nl}/\partial t$, becomes large (see Supplemental material), yielding a large current $I_{c}^{(nl)}$, as depicted in Fig.\ \ref{IP_different_BSR}(c).} This explains the vanishing hysteresis, noting that the geometric constant capacitance $C_0$ is very small. Otherwise, if the constant capacitance is larger, one may expect a significant hysteresis, knowing that increasingly higher rates would produce larger and larger hysteretic effects. This particular aspect is shown in the inset of Fig.\ \ref{IP_different_BSR}(a), where a very large dielectric constant of the perovskite layer, $\epsilon_r=10^5$, was considered for comparison.
Here one should take into account CH$_3$NH$_3$PbI$_3$ has demonstrated a large static dielectric constant of $\sim$70 \cite{lin} or higher, due to electronic, ionic, and molecular dipole 
contributions from the organic cation \cite{brivio} and, in other studies, a giant dielectric constant has been reported \cite{perez}. In this case, while the geometrical capacitance increases from $\mu$F to mF range, the hysteresis becomes more pronounced at higher BSRs, in particular at 1000 mV/s. By contrast, at low scan rates, for $\alpha=20$ mV/s, there is practically no difference in the hysteresis compared to the reference curve, obtained for $\epsilon=100$.

The experimental data upon which the sequence of simulated J-V curves were calibrated is presented in Fig.\ \ref{IP_different_BSR}\cred{(d)}. It is the same sample analyzed in Section \ref{pre-poling}, and the simulation parameters are also the same, except for a slightly smaller current due to aging, $I_{ph}=0.43$mA, and $P_0=5 P_\infty$ reflecting the smaller initial preconditioning voltage $V_0=1$V. Under these conditions the experimental J-V curve for $\alpha=20$ mV/s is well reproduced by the DEM results. 
Then, comparing Figs.\ \ref{IP_different_BSR}(a) and \cred{(d)}, the behavior obtained by changing the BSR is very similar, albeit at higher rates, e.g. 1000 mV/s, the simulated hysteresis is smaller. However, increasing the dielectric constant, the hysteresis becomes larger, qualitatively and quantitatively similar with experimental data. At even higher rates, it is expected that $\epsilon_r$ decreases, as reported also in Ref.\ \cite{perez}, diminishing the geometrical capacitance and the associated hysteresis.

Another similarity between our results and the reported experimental data \cite{tress} is the trend exhibited by the short circuit current: it increases with the bias scan rate. This again may be connected with the non-linear polarization relaxation $\partial P_{nl}/\partial t$, which is faster for higher BSRs, leading to larger currents, while $\partial P_{nl}/\partial V$ becomes small and $P_{nl}$ is similar for both reverse and forward scans, which is consistent with a reduced hysteretic effect.

\subsection{Different relaxation time scales}

\begin{figure}[t]
\centering
\includegraphics[width=8.cm]{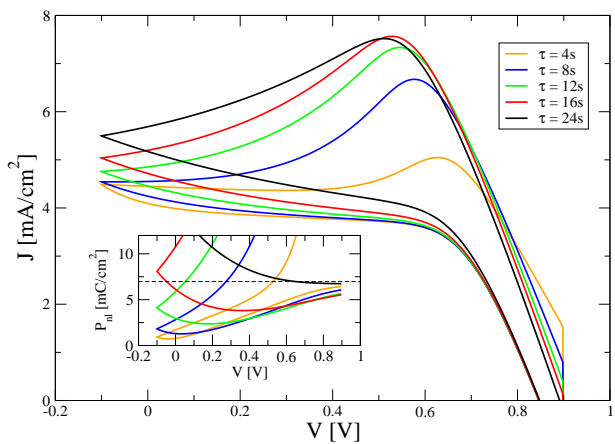}
\caption{Dynamic J-V characteristics and polarization under reverse-forward bias scan, for different $\tau$ values. The inset indicates the slower decay of the $P_{nl}$ polarization term for larger relaxation time scales.}
\label{IP_different_tau}
\end{figure}

Taking as reference the parameters used in section \ref{pre-poling}, we present in Fig.\ \ref{IP_different_tau} the J-V characteristics for different relaxation times in Eq.\ (\ref{dPdt}), $\tau=4, 8, 12, 16, 24$ s. Lowering $\tau$ the overshoot observed in the reverse scan is displaced towards higher biases, which is consistent with the fact the relaxation occurs faster and, in addition, at $V_{oc}$ there is a higher jump in the current. For the smallest characteristic time $\tau=4$ s, the largest difference between the forward and reverse scans is found at large bias, $V>0.5$ V, while at smaller biases the hysteresis is smaller.  
For larger relaxation times, $\tau=24$ s, the overshoot in reverse is flattened and the hysteresis begins to decrease. 
The polarization depicted in the inset of Fig.\ \ref{IP_different_tau} presents a more rapid decrease for smaller relaxation times. 
This phenomenology may be consistently linked to the previous discussion about the changing the bias scan rate, and implicitly the time during which the J-V scans are performed.
On the other hand, as briefly noted in section \ref{pre-poling}, the current overshoot is displaced towards smaller biases, which may be correlated with an enhanced relaxation timescale when larger pre-poling is applied. 
Multi-exponential signal decays have also been reported \cite{tress,roiati,lee}, ranging from microseconds to seconds, being attributed to different phenomena of electronic and non-electronic nature.
This provides motivation for further exploration of multiple time scales in the DEM, correlated with microscopic models.

\section{Conclusions}
\label{conclusions}

A dynamic electrical model was introduced to describe hysteresis effects observed in halide perovskite based solar cells. Using a particularly simple ansatz for the polarization relaxation one may reproduce detailed features of experimental J-V characteristics, both qualitatively and quantitatively. For one, the maximum typically observed in the reverse bias scan is attributed to an initial over-polarization effect. On the other hand, initial under-polarization introduces a crossing of the reverse and forward scans, which almost overlap at large biases, with no maximum in reverse characteristics. By changing the bias scan rate, our model is able to reproduce details of the experimental data such as the short circuit current and the reverse current overshoot. Furthermore, in this context, geometrical capacitance effects are also discussed.
Additional analysis is performed to quantify the effects related to the magnitude of the characteristic relaxation time scales, which are further correlated with pre-poling experiments.  
By analyzing the perovskite based solar cells under non-stationary bias conditions one may extract further knowledge regarding the device operation and transport mechanisms.
The proposed dynamic electrical model accurately describes the measured dynamic J-V characteristics 
and therefore it may constitute a valuable tool in the calibration of microscopic models relevant for solar cell operation.\\

{\bf Acknowledgements} \\

The research leading to these results has received funding from EEA Financial Mechanism 2009-2014 under the project contract no 8SEE/30.06.2014.


\section*{References}


\newpage

\onecolumn

\section*{Supplemental material}

\begin{figure}[h]
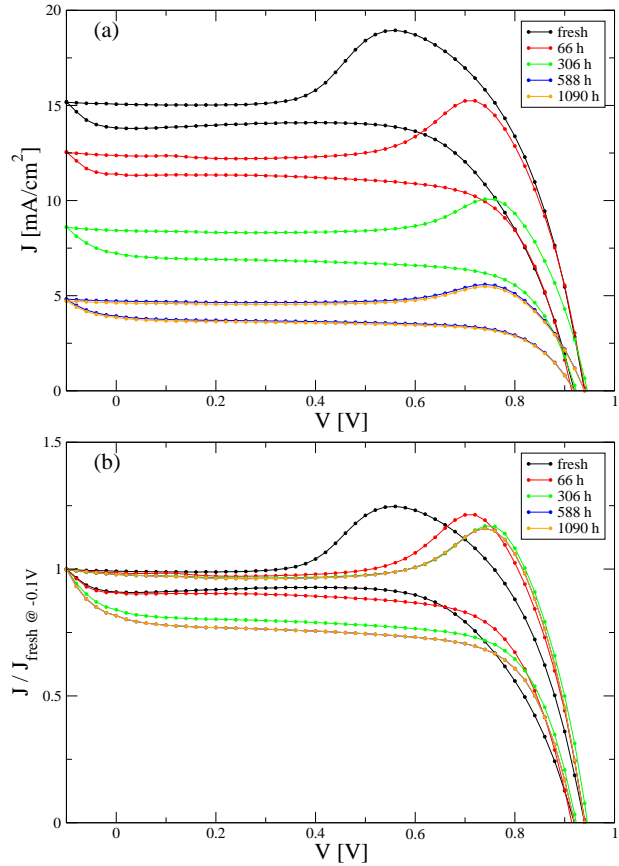

\centering
\includegraphics[width=8.cm]{S1a}\\
\includegraphics[width=8.cm]{S1b}
\caption{(a) Aging of the J-V characteristics during time: measurements on the fresh cells and after 66, 306 588, 1090 hours. The scan rate was 20 mV/s.
(b) Scaled J-V characteristics with the current of the fresh cell measured at -0.1 V.}
\label{aging}
\end{figure}

{\bf A.} The aging of solar cells was observed and monitored during time. The samples were kept in the dark and inert atmosphere conditions. 
Several J-V characteristics measured over a period of 45 days are shown in Fig.\ \ref{aging}(a). They all present hysteresis irrespective of age. Reverse-forward bias scans were performed, starting from the pre-poling bias of 1V.
The calculated maximum PCE is evidently different for the reverse and forward bias scans. The highest PCE values are obtained for the fresh cell, 7.95\% (forward) and 11.87\% (reverse). Subsequent measurements indicate lower values for each type of bias scan. It is however important to note that a significant degradation occurs in the first $\sim$ 588 hours, while further on the PCE is practically stabilized at $\sim$ 4.2\%. The maximum PCEs obtained are indicated in Table \ref{PCEs}.
In Fig.\ \ref{aging}(b) the scaled J-V characteristics show the hysteresis of the normalized currents increases over time. The currents are scaled with the value obtained for the fresh cell measured at -0.1V. Furthermore one should note the rather precise scaling of the aged cells over the entire bias range.
 
\begin{table}[h]
\caption{\label{PCEs} Maximum PCEs obtained in the forward (F) and reverse (R) bias scans: }

\begin{center}
\begin{tabular}{|l|r|r|r|r|r|}
\hline \hline
Time  & \hspace*{0.75cm}fresh & \hspace*{0.75cm}66 h & \hspace*{0.75cm}306 h & \hspace*{0.75cm}588 h & \hspace*{0.75cm}1090 h \cr
\hline \hline
F-scan & \hspace*{0.75cm}7.95 & \hspace*{0.75cm}6.69 & \hspace*{0.75cm}4.43 & \hspace*{0.75cm}2.44 & \hspace*{0.75cm}2.39 \cr
\hline
R-scan & \hspace*{0.75cm}11.87 & \hspace*{0.75cm}11.08 & \hspace*{0.75cm}7.66 & \hspace*{0.75cm}4.21 & \hspace*{0.75cm}4.15 \cr
\hline \hline
\end{tabular}
\end{center}
\end{table}

\begin{figure}[h]
\centering
\includegraphics[width=8.cm]{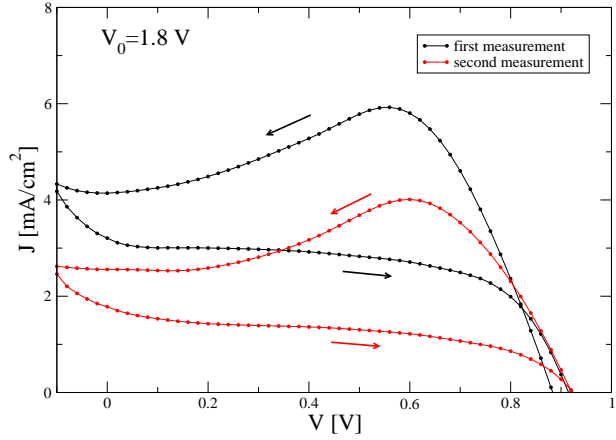}
\caption{Degradation of the J-V characteristics after pre-poling at the high bias $V_0 = 1.8$ V, measured at a BSR of 20 mV/s.}
\label{damage}
\end{figure}

{\bf B.} Concerning the damage of the device, our experiments consistently showed that after bias stressing of the cell the overall shape of the I-V curves always is preserved, while the current suffers a significant decrease which also affects the PCE. The degradation of the cells after consecutive measurements was observed regardless of the applied initial bias value. For exemplification we have included data recorded for a pre-poling bias of 1.8 V, shown in Fig.\ \ref{damage} in two consecutive measurements performed on an aged cell.\\

\begin{figure}[h]
\centering
\includegraphics[width=8.cm]{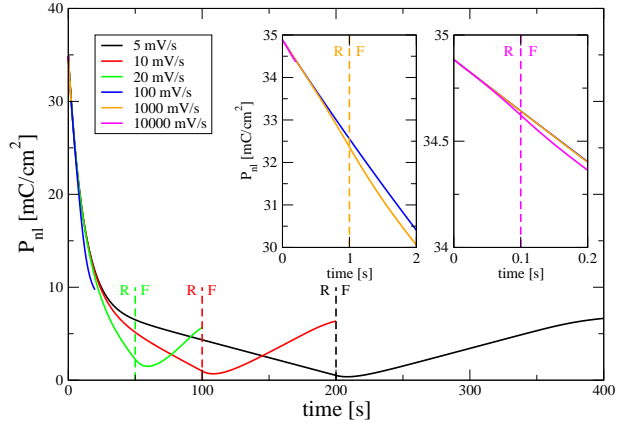}
\caption{The non-linear polarization $P_{nl}$ vs. time, for different BSRs. The rapid variation of $P_{nl}$ for the largest BSRs considered, 1000 mV/s and 10000 mV/s, are presented in detail in the two insets. The vertical dashed lines mark the reverse/forward (R/F) regimes.
}
\label{Pt_Icnl}
\end{figure}

{\bf C.} We discuss in more detail the increase of the short-circuit current at high bias scan rates (BSRs), which
was observed in our experimental data as well as in Ref. [18]. At small biases (V $<$ $\sim$0.4V), in
 the case of over-polarized samples, a significant current component is represented by $J_c^{(nl)}$,
 which stems from the time variation of the non-linear polarization $P_{nl}$. This can be seen e.g.
in Fig. 3, where $J \approx J_{ph} - J_c^{(nl)}$ (V $<$ $\sim$0.4V). The time dependency of $P_{nl}$ for different bias scan rates is depicted in Fig.\ \ref{Pt_Icnl}. Given the BSRs of 5, 10, 20, 100, 1000 and 10000 mV/s, the time intervals for the reverse-forward bias scans in the bias range [-0.1,0.9] V are correspondingly 400, 200, 100, 20, 2 and 0.2 seconds. Due to the relatively large relaxation time ($\tau=11.5$s) and the short amount of time spent in the negative bias range,
the polarization is always positive. For low BSRs (5 - 20 mV/s) a minimum of polarization is observed around V = -0.1V bias point, with a retardation of the order of $\tau$. In this case, a rapid decrease (in absolute value) of the $J_c^{(nl)}$ current is observed on reverse scans in Fig.\ 5(c), followed by a sign change, as $P_{nl}$ reaches its minimum.
Increasing further the BSR, $P_{nl}$ decreases monotonically, with higher slopes, as one can see in the two insets. For the two largest BSRs, 1000 and 10000 mV/s, the decrease of
$P_{nl,\infty}(U_c(t))$ on the reverse scan is so rapid with respect to the relaxation time $\tau$, so that $P_{nl}$ becomes a concave function of time.
On the forward scan $P_{nl}(t)$ becomes a convex function as 
$P_{nl,\infty}(U_c(t))$ increases, the two regimes being separated by an inflection point. This explains the increase of $J_c^{(nl)}\propto\partial P_{nl}/\partial t$ (absolute value) on reverse scan at low biases in Fig.\ 5(c), for the BSRs of 1000 and 10000 mV/s, and the opposite is found for the forward scan.

\end{document}